\documentclass[prd,twocolumn,superscriptaddress,altaffilletter,nofootinbib]{revtex4}

\usepackage{cancel}
\usepackage{graphicx}
\usepackage{amsmath}
\usepackage{amssymb}
\usepackage{graphicx,epsfig}
\newtheorem{theorem}{Theorem}
\newtheorem{corollary}{Corollary}[theorem]
\newtheorem{lemma}[theorem]{Lemma}

\newcommand{\be}{\begin{equation}}
\newcommand{\ee}{\end{equation}}
\newcommand{\bea}{\begin{eqnarray}}
\newcommand{\eea}{\end{eqnarray}}
\newcommand{\der}{\partial}
\newcommand{\vphi}{\varphi}
\newcommand{\bet}{\begin{theorem}}
\newcommand{\eet}{\end{theorem}}
\newcommand{\bec}{\begin{corollary}}
\newcommand{\eec}{\end{corollary}}

\begin{document}



\title{Phenomenological signatures of gauge invariant theories of gravity with vectorial nonmetricity}



\author{Israel Quiros}\affiliation{Dpto. Ingenier\'ia Civil, Divisi\'on de Ingenier\'ia, Universidad de Guanajuato, Gto., M\'exico.}



\begin{abstract}
In this paper we discuss on the phenomenological footprints of gauge invariant theories of gravity where the gravitational effects are due not only to spacetime curvature, but also to vectorial nonmetricity. We explore the possibility that vectorial nonmetricity and gauge symmetry may survive after $SU(2)\times U(1)$ (electroweak) symmetry breaking, so that these may have impact on the explanation of certain cosmological puzzles, such as the nature of the dark matter and of the dark energy. We show that this is possible only for theories with gradient nonmetricity, i. e., when the vectorial nonmetricity amounts to a gradient of a scalar. The possibility that vectorial nonmetricity may have played a role in the quantum epoch is not ruled out. We also present an alternative interpretation of gauge invariance of theories with vectorial nonmetricity which we call as ``many-worlds'' interpretation due to its overall similitude with the known interpretation of quantum physics. 
\end{abstract}



\maketitle


\section{Introduction}

 
Weyl geometry \cite{weyl-1917}, the theoretical framework where gauge symmetry was introduced for the first time, played an important role in the early search for alternatives on unification of the fundamental interactions \cite{london-1927, dirac-1973, utiyama-1973, adler-book, maeder-1978, smolin-1979, cheng-1988, cheng-arxiv, perlick-1991, many-weyl-book}. It represented an interesting generalization of Riemann geometry, nevertheless, discussions on the occurrence of the second clock effect (SCE) \cite{tucker, 2clock-1, 2clock-2, 2clock-4, 2clock-5, 2clock-tomi, 2clock-delhom, delhom-coupling, hobson, quiros-2022, tomi-replay, hobson-replay}, ruled it out as phenomenologically nonviable. Recently nonmetricity theories, where the covariant derivative of the metric does not vanish \cite{delhom-2019},

\bea \nabla_\alpha g_{\mu\nu}=-Q_{\alpha\mu\nu},\label{gen-nm}\eea with $Q_{\alpha\mu\nu}$ -- the nonmetricity tensor, have played an interesting role in the search for alternative explanations to fundamental questions of current interest. The recent resurrection of nonmetricity theories is mainly focused in the so called teleparallel \cite{tele-rev-capoz, tele-ferraro, tele-maluf, tele-soti, tele-brazil, tele-baha, tele-nester, tele-coley, tele-intro} and, specially, the symmetric teleparallel theories \cite{adak-2006, adak-2006-1, adak-2013, beltran-plb-2016, javr-prd-2018, vilson-prd-2018, formiga-2019, adak-arxiv, obukhov, beltran-j-universe} and their cosmological applications \cite{lavinia-review, beltran-prd-2020, saridakis-prd-2020, lazkoz-prd-2019, lazkoz-prd-2021, sanjay-prd-2020, sanjay-2021}. Despite that Weyl gauge invariance is an intrinsic symmetry of nonmetricity geometry, in the bulk of these papers, gauge symmetry is ignored. Only in a few bibliographic references nonmetricity is investigated from the point of view of gauge symmetry \cite{saridakis-prd-2020, quiros-2022, adak-2022}. 

Gauge symmetry is one of the most important properties of nonmetricity geometry due to covariance of \eqref{gen-nm} under the following Weyl gauge transformations \cite{delhom-2019, saridakis-prd-2020}:

\bea g_{\mu\nu}\rightarrow\Omega^2g_{\mu\nu},\;Q^\alpha_{\;\;\mu\nu}\rightarrow Q^\alpha_{\;\;\mu\nu}-2\der^\alpha\ln\Omega\,g_{\mu\nu},\label{gen-gauge-t}\eea where the positive smooth function $\Omega=\Omega(x)$, is the conformal factor. In \eqref{gen-gauge-t} the conformal transformation of the metric does not represent a diffeomorphism or, properly, a conformal isometry, i. e., the spacetime coincidences or events (as well as the spacetime coordinates that label the events,) are not modified by the conformal transformations. Here we shall focus in a class of gauge invariant gravitational theories with vectorial nonmetricity, i. e. when in Eq. \eqref{gen-nm} we set $Q_{\alpha\mu\nu}=Q_\alpha g_{\mu\nu}$. We shall not consider the teleparallel condition (vanishing overall Riemannian curvature) so that the teleparallel and the symmetric teleparallel theories, as well as their cosmological applications, fall beyond of the scope of this paper.

In the case of a gauge invariant theory of gravity over background spacetimes with nonmetricity vector $Q_\mu$, gauge invariance means that the gravitational laws are not affected by simultaneous conformal transformations of the metric and gauge transformations of the nonmetricity vector:

\bea g_{\mu\nu}\rightarrow\Omega^2g_{\mu\nu},\;Q_\mu\rightarrow Q_\mu-2\der_\mu\ln\Omega,\label{gauge-t}\eea together with appropriate transformations of the remaining fields according to their conformal weight $w$: $\Psi_a\rightarrow\Omega^w\Psi_a$. In what follows we shall call the transformations \eqref{gauge-t} either as Weyl gauge transformations or, simply, as gauge transformations.

The standard Weyl gauge invariant picture can be explained as follows. $SU(2)\times U(1)$ symmetry breaking may be associated with the Higgs Lagrangian,

\bea {\cal L}_H=\frac{1}{2}|D_gH|^2+\frac{\lambda'}{2}\left(|H|^2-v^2_0\right)^2,\label{higgs-lag}\eea where $v_0$ is the constant electroweak (EW) mass parameter, $H$ is the Higgs doublet, and we use the following notation: $|H|^2\equiv H^\dag H$, $|D_gH|^2\equiv g^{\mu\nu}(D^g_\mu H)^\dag(D^g_\nu H)$, and

\bea D^g_\mu H=\left(\der_\mu+\frac{i}{2}g W^k_\mu\sigma^k+\frac{i}{2}g' B_\mu\right)H,\label{ew-gauge-der}\eea with $W^k_\mu$ -- the $SU(2)$ bosons, $B_\mu$ -- the $U(1)$ boson, $(g,g')$ -- gauge couplings and $\sigma^k$ are the Pauli matrices. After $SU(2)\times U(1)$ symmetry breaking the standard model (SM) particles acquire constant masses which break the gauge symmetry \cite{deser-1970}. It seems that there is no room for gauge symmetry and, correspondingly, for nonmetricity in the phenomenology of the SM after $SU(2)\times U(1)$ symmetry breaking. In particular, nonmetricity may not account neither for the dark matter (DM) -- as it has been repeatedly claimed \cite{cheng-1988, cheng-arxiv, mannheim-1989, mannheim-2006, harko-prd-2023} -- nor for the dark energy (DE), since the DM-dominated and the present DE-dominated stages of the cosmic evolution take place long after matter-radiation decoupling. Typically, after EW and Weyl gauge symmetry breaking, Riemann geometry structure of background space is recovered since the nonmetricity field acquires a large mass $\propto M_\text{pl}$, rendering the nonmetricity effects suppressed by this scale \cite{ghilen-1} (se also \cite{ghilen-2023, ghilen-2, ghilen-3, ghilen-4}.) We recover general relativity (GR) over Riemann space as the theory describing the gravitational interactions of matter since the symmetry breaking event and up to the present, so that DM/DE phenomenology can not be explained.


One of the goals of the present paper is to search for possible cosmological consequences of a class of gauge invariant gravitational theories over background space with nonmetricity vector $Q_\mu$, which are driven by second-order equations of motion (EOM) exclusively. This means that we avoid any Ostrogradsky ghosts, among other instabilities. The theories in this class are given by the following gravitational action over Weyl geometry space:

\bea S_g=\alpha\int d^4x\sqrt{-g}\left[\phi^2R+\omega(\der^*\phi)^2-\frac{\lambda}{4}\phi^4-\frac{\beta^2}{2}Q^2\right],\label{g-action'}\eea where the curvature scalar of Weyl space $R$ is given by \eqref{gen-curv-sc}, $\omega$, $\lambda$ and $\beta^2$ are free coupling parameters, $\der^*_\mu\phi\equiv\der_\mu\phi-Q_\mu\phi/2$ and we have introduced the shorthand notation $Q^2\equiv Q_{\mu\nu}Q^{\mu\nu}$ (the nonmetricity field strength is defined as $Q_{\mu\nu}:=2\der_{[\mu}Q_{\nu]}$.) In what follows, without loss of generality, we set $\alpha=1/2$. 

If this class of theories is supposed to have cosmological consequences, these should be able to describe the phenomenology after EW symmetry breaking. Hence, the $SU(2)\times U(1)$ symmetry breaking Lagrangian \eqref{higgs-lag} has to be modified in such a way that Weyl gauge symmetry is preserved. The required modification amounts to lifting the mass parameter $v_0$ to a point dependent field \cite{bars-2014, quiros-2014}: $v_0\rightarrow v(x)$, such that under \eqref{gen-gauge-t}, $v^2\rightarrow\Omega^{-2}v^2$. Besides, the EW gauge covariant derivative in Eq. \eqref{ew-gauge-der} is to be replaced as well: $D^g_\mu\rightarrow D^g_\mu-Q_\mu/2$, so that, under the gauge transformations \eqref{gen-gauge-t}, $D^g_\mu H\rightarrow\Omega^{-1}D^g_\mu H$ $\Rightarrow\;|D_gH|^2\rightarrow\Omega^{-4}|D_gH|^2$. Lifting the mass parameter to a point dependent field $v(x)$ leads to the masses acquired by the particles of the SM after EW symmetry breaking, being point dependent quantities as well:\footnote{Point dependent masses which transform under the conformal transformation of the metric as $m\rightarrow\Omega^{-1}m$, are considered by Dicke in his paper \cite{dicke-1962} and in subsequent papers on conformal transformations in scalar-tensor theories (STT) \cite{cf-stt-1, cf-stt-2, cf-stt-3, cf-stt-4, cf-stt-5, cf-stt-6, cf-stt-7, quiros-grg-2013}.} $m=m(x)$. Under \eqref{gen-gauge-t} the mass $m$ of given particle transforms like $m\rightarrow\Omega^{-1}m$. This means that the presence of nonvanishing masses does not represent a risk for gauge symmetry.

One of the main results of our study is presented in the form of a lemma stating that (i) only matter fields with vanishing trace of the stress-energy tensor (SET) couple to gravity in models with gravitational action \eqref{g-action'} and that (ii) these matter fields follow null geodesics of Riemann geometry. A corollary of this lemma is that radiation and massless fields -- the only matter fields which interact with gravity in theories within the class given by \eqref{g-action'} -- do not interact with the nonmetricity vector $Q_\mu$. Although there are demonstrations in the bibliography proving that the Weyl gauge vector $Q_\mu$ does not interact with massless fermions and with other massless gauge fields, but only with gravitation \cite{cheng-1988, cheng-arxiv}, the statement and proof of our lemma is more general and does not require of specific Lagrangian for matter fields. Besides former demonstrations seem to incorrectly suggest that the nonmetricity vector $Q_\mu$ can take account of the dark matter component of the cosmic budget. Such a conclusion is incorrect since $Q_\mu$ (jointly with the metric field) is a carrier of the gravitational interactions of matter. Hence, the fact that radiation and massless fields do not interact with $Q_\mu$, which represents the non Riemannian contribution to gravity, means that the nonmetricity vector may be removed without physical consequences. All of this is clearly established at once by our more general demonstration in this paper. We also demonstrate that the only gauge invariant gravitational theory with nonvanishing nonmetricity leading to second-order EOM, which admits coupling of all of the SM fields, no matter whether massless or not, is the one with gradient nonmetricity, i. e., when $Q_\mu=\der_\mu\vphi$, where $\vphi$ is a gauge scalar. 


Another goal of this paper is to discuss, from perhaps a non conventional perspective, on a very important consequence of gauge invariance: gauge freedom and gauge fixing. Within the gravitational context gauge freedom has a different interpretation than, for instance, electromagnetic (EM) $U(1)$ gauge freedom. In order to understand this (apparently trivial) statement let us briefly explain the implications of the EM $U(1)$ gauge symmetry. From the operational point of view, EM $U(1)$ gauge fixing entails a mathematical constraint on (the derivatives of) the EM potential $A_\mu$, allowing elimination of one redundant degree of freedom (DOF) and simplification of subsequent computations. The physical interpretation of gauge symmetry when the EM field is coupled to matter (for instance to fermions,) is nicely discussed in section 6 of \cite{tong-2007}. In this case the EM Lagrangian density reads,

\bea {\cal L}_\text{EM}=\bar\psi(i\cancel D-m)\psi+\frac{1}{4}F_{\mu\nu}F^{\mu\nu},\label{em-lag}\eea where $\psi$ is the fermion's spinor while $\bar\psi:=\psi^\dag\gamma^0$ represents its Dirac adjoint, $\gamma^\mu$ are the Dirac gamma matrices, $m$ stands for the mass of the spinor field, $\cancel D:=\gamma^\mu(\der_\mu+ie A_\mu)$ and $F_{\mu\nu}:=\der_\mu A_\nu-\der_\nu A_\mu$ are the coordinate components of the EM field strength. The EM Lagrangian \eqref{em-lag}, as well as the derived EOM, are invariant under the $U(1)$ gauge transformations:

\bea \psi\rightarrow e^{-ie\lambda(x)}\psi,\;\bar\psi\rightarrow e^{ie\lambda(x)}\bar\psi,\;A_\mu\rightarrow A_\mu+\der_\mu\lambda(x),\label{u1-gauge-t}\eea where $\lambda(x)$ can be any function. We have an infinite set of possible descriptions $(\psi,\bar\psi,A_\mu)$ due to the freedom in the choice of the function $\lambda$. Thanks to the fact that the scalar density $\rho_\psi\propto\bar\psi\psi$, which carries the relevant information about the quantum state of the fermion, is not affected by phase shifts $\sim\lambda(x)$, any two states, picked out by two different choices $\lambda_1(x)$ and $\lambda_2(x)$, are to be identified. This means that a specific choice of gauge carries no physical consequences at all.

In the case of Weyl gauge symmetry, fermion's spinor and its Dirac adjoint $\bar\psi$ transform in the same way under \eqref{gauge-t}, i. e. both share the same conformal weight: $w(\psi)=w(\bar\psi)=-3/2$. This means that, under the conformal transformation of the metric in \eqref{gauge-t}, the scalar density $\rho_\psi\propto\bar\psi\psi$, transforms like $\rho_\psi\rightarrow\Omega^{-3}\rho_\psi$. Hence, contrary to EM $U(1)$ gauge symmetry, the fermion's scalar density is transformed by the gauge transformations, so that we can not identify two different states of the fermion corresponding to two different choices of the conformal factor: $\Omega_1(x)$ and $\Omega_2(x)$, respectively. Besides, conformal transformations of the metric in \eqref{gauge-t} link two different metrics, i. e., two different ways of measuring distances in spacetime. Each one of the conformally related metrics leads to different curvature properties encoded in the curvature tensors: Riemann-Christoffel curvature tensor and its contractions. Hence, a gauge invariant theory of gravity is not a single theory but a conformal equivalence class of them. Conformal transformations with different $\Omega(x)$-s link the different theories in the equivalence class. In this context gauge fixing amounts to choosing a specific theory of the gravitational interactions of matter in the equivalence class. Therefore, unlike EM gauge choice which carries no physical consequences, the choice of gauge within a gauge invariant gravitational theory has far reaching physical consequences. 

Here we develop an alternative understanding of gauge fixing in gravitational theories which, despite obvious differences, bears resemblance with the many-worlds interpretation of quantum physics. According to our alternative interpretation each gauge choice picks out one possible theory of gravity in the conformal equivalence class. Not every gauge choice, although representing a potential description of our universe, gives a phenomenologically viable description. Besides, we can determine the gauge where we and the rest of the matter fields in the universe live in: this is the one which better describes the existing amount of observational and experimental evidence at once.


This paper has been organized in the following way. In section \ref{sect-basic} we expose the notation as well as the basic mathematical knowledge required to understand the main text. In section \ref{sect-ginv} we derive the EOM as well as the main properties of the class of gauge invariant gravitational theories given by the action \eqref{g-action'}. One of the main results of the present paper: (i) that only matter fields with vanishing SET trace couple to gravity and (ii) that these fields follow null geodesics of Riemann space, is presented in the form of lemma 1 -- along with its proof -- in section \ref{sect-lemma-1}. In section \ref{sect-low-energy} we discuss on the low-energy spectrum of the gauge invariant gravitational theory \eqref{g-action'}. An innovative (and perhaps controversial) aspect of gauge symmetry proposed in the present work: the many-worlds approach to gauge freedom, is discussed in section \ref{sect-g-freedom}. This approach is illustrated in section \ref{sect-example}, through a cosmological example. In section \ref{sect-wig} we demonstrate that gradient nonmetricity is the only possibility left for gauge invariance to play a role in phenomenology after EW symmetry breaking. Discussion of the main results of this research, as well as brief conclusions, are given in section \ref{sect-discu}. In this section we compare the results of the present investigation with several formerly published results.



\section{Background and conventions}\label{sect-basic} 



Unless otherwise stated, here we use natural units where $\hbar=c=1$ and the following signature of the metric is chosen: $(-+++)$. Greek indices run over spacetime $\alpha,\beta,...,\mu,...=0,1,2,3$, while latin indices $i,j,k...=1,2,3$ run over three-dimensional space. 


Weyl geometry space, denoted here by $\tilde W_4$, is defined as the class of four-dimensional (torsionless) manifolds ${\cal M}_4$ that are paracompact, Hausdorff, connected $C^\infty$, endowed with a locally Lorentzian metric $\bf g$ that obeys the vectorial nonmetricity condition:\footnote{When the generalized nonmetricity condition \eqref{gen-nm} is satisfied, the resulting space is denoted by $W_4$ and it is called as generalized Weyl space \cite{quiros-2022}. Standard Riemann space, which is characterized by vanishing nonmetricity: $\nabla_\alpha g_{\mu\nu}=0$, is denoted by $V_4$.}

\bea \nabla_\alpha g_{\mu\nu}=-Q_\alpha g_{\mu\nu},\label{vect-nm}\eea where $Q_\alpha$ is the Weyl gauge vector and the covariant derivative $\nabla_\mu$ is defined with respect to the torsion-free affine connection of the manifold: 

\bea \Gamma^\alpha_{\;\;\mu\nu}=\{^\alpha_{\mu\nu}\}+L^\alpha_{\;\;\mu\nu},\label{gen-aff-c}\eea where 

\bea \{^\alpha_{\mu\nu}\}:=\frac{1}{2}g^{\alpha\lambda}\left(\der_\nu g_{\mu\lambda}+\der_\mu g_{\nu\lambda}-\der_\lambda g_{\mu\nu}\right),\label{lc-aff-c}\eea is the Levi-Civita (LC) connection, while $L^\alpha_{\;\;\mu\nu}:=(Q_\mu\delta^\alpha_\nu+Q_\nu\delta^\alpha_\mu-Q^\alpha g_{\mu\nu})/2,$ is the disformation tensor. The Weyl gauge vector $Q_\alpha$ measures how much the length of given timelike vector varies during parallel transport.

In this paper we call as ``generalized curvature tensor'' of $\tilde W_4$ spacetime, the curvature of the connection, whose coordinate components are,

\bea R^\alpha_{\;\;\sigma\mu\nu}:=\der_\mu\Gamma^\alpha_{\;\;\nu\sigma}-\der_\nu\Gamma^\alpha_{\;\;\mu\sigma}+\Gamma^\alpha_{\;\;\mu\lambda}\Gamma^\lambda_{\;\;\nu\sigma}-\Gamma^\alpha_{\;\;\nu\lambda}\Gamma^\lambda_{\;\;\mu\sigma},\label{gen-curv-t}\eea or, if take into account the decomposition \eqref{gen-aff-c}:

\bea &&R^\alpha_{\;\;\sigma\mu\nu}=\hat R^\alpha_{\;\;\sigma\mu\nu}+\hat\nabla_\mu L^\alpha_{\;\;\nu\sigma}-\hat\nabla_\nu L^\alpha_{\;\;\mu\sigma}\nonumber\\
&&\;\;\;\;\;\;\;\;\;\;\;\;\;\;\;\;\;\;\;\;\;\;\;\;\;+L^\alpha_{\;\;\mu\lambda}L^\lambda_{\;\;\nu\sigma}-L^\alpha_{\;\;\nu\lambda}L^\lambda_{\;\;\mu\sigma},\label{gen-curv-t-1}\eea where $\hat R^\alpha_{\;\;\sigma\mu\nu}$ is the Riemann-Christoffel or LC curvature tensor,\footnote{In this paper quantities and operators with a hat are defined with respect to the LC connection \eqref{lc-aff-c}.}

\bea \hat R^\alpha_{\;\;\sigma\mu\nu}:=\der_\mu\{^\alpha_{\nu\sigma}\}-\der_\nu\{^\alpha_{\mu\sigma}\}+\{^\alpha_{\mu\lambda}\}\{^\lambda_{\nu\sigma}\}-\{^\alpha_{\nu\lambda}\}\{^\lambda_{\mu\sigma}\},\label{lc-curv-t}\eea and $\hat\nabla_\alpha$ is the LC covariant derivative. Besides, the LC Ricci tensor $\hat R_{\mu\nu}=\hat R^\lambda_{\;\;\mu\lambda\nu}$ and LC curvature scalar read:

\bea &&\hat R_{\mu\nu}=\der_\lambda\{^\lambda_{\nu\mu}\}-\der_\nu\{^\lambda_{\lambda\mu}\}+\{^\lambda_{\lambda\kappa}\}\{^\kappa_{\nu\mu}\}-\{^\lambda_{\nu\kappa}\}\{^\kappa_{\lambda\mu}\},\nonumber\\
&&\hat R=g^{\mu\nu}\hat R_{\mu\nu},\label{lc-curv-sc}\eea respectively. We call $R^\alpha_{\;\;\sigma\mu\nu}$ as generalized curvature tensor because it is contributed both by LC curvature $\hat R^\alpha_{\;\;\sigma\mu\nu}$, and by nonmetricity through disformation $L^\alpha_{\;\;\mu\nu}$. We have that,

\bea &&R_{\mu\nu}=\hat R_{\mu\nu}+\hat\nabla_\lambda L^\lambda_{\;\;\mu\nu}-\hat\nabla_\nu L^\lambda_{\;\;\lambda\mu}\nonumber\\
&&\;\;\;\;\;\;\;\;\;\;\;\;\;\;\;\;\;\;+L^\lambda_{\;\;\lambda\kappa}L^\kappa_{\;\;\mu\nu}-L^\lambda_{\;\;\nu\kappa}L^\kappa_{\;\;\lambda\mu},\label{gen-ricci-t}\\
&&R=\hat R-\frac{3}{2}Q_\mu Q^\mu-3\hat\nabla_\mu Q^\mu.\label{gen-curv-sc}\eea  

The generalized curvature tensor $R^\alpha_{\;\;\sigma\mu\nu}$ has various contractions. In order to show these contractions let us write Eq. \eqref{gen-curv-t-1} in the following form:

\bea &&R_{\alpha\sigma\mu\nu}=\hat R_{\alpha\sigma\mu\nu}+\hat\nabla_\mu L_{\alpha\nu\sigma}-\hat\nabla_\nu L_{\alpha\mu\sigma}\nonumber\\
&&\;\;\;\;\;\;\;\;\;\;\;\;\;\;\;\;\;\;\;\;\;\;\;\;\;+L_{\alpha\mu\lambda}L^\lambda_{\;\;\nu\sigma}-L_{\alpha\nu\lambda}L^\lambda_{\;\;\mu\sigma}.\label{gen-curv-tensor}\eea The various linearly independent contractions of the generalized curvature tensor are,

\bea &&R_{\mu\nu}:=g^{\lambda\kappa}R_{\lambda\mu\kappa\nu},\;\tilde R_{\mu\nu}:=g^{\lambda\kappa}R_{\mu\lambda\nu\kappa},\nonumber\\
&&R=g^{\mu\nu}R_{\mu\nu}=g^{\mu\nu}\tilde R_{\mu\nu}.\label{var-contract}\eea The first two of these amount to, 

\bea &&R_{\mu\nu}=\hat R_{\mu\nu}-\frac{1}{2}\left(Q_\lambda Q^\lambda+\hat\nabla_\lambda Q^\lambda\right)g_{\mu\nu}+\frac{1}{2}Q_\mu Q_\nu\nonumber\\
&&\;\;\;\;\;\;\;\;\;\;\;\;\;\;\;\;\;\;\;-\hat\nabla_\nu Q_\mu+\frac{1}{2}\left(\hat\nabla_\mu Q_\nu-\hat\nabla_\nu Q_\mu\right),\label{1-ricci-t}\eea and to,

\bea &&\tilde R_{\mu\nu}=\hat R_{\mu\nu}-\frac{1}{2}\left(Q_\lambda Q^\lambda+\hat\nabla_\lambda Q^\lambda\right)g_{\mu\nu}+\frac{1}{2}Q_\mu Q_\nu\nonumber\\
&&\;\;\;\;\;\;\;\;\;\;\;\;\;\;\;\;\;\;\;-\frac{1}{2}\left(\hat\nabla_\mu Q_\nu+\hat\nabla_\nu Q_\mu\right),\label{2-ricci-t}\eea respectively. We shall call $R_{\mu\nu}$ as first Ricci tensor while $\tilde R_{\mu\nu}$ we shall call as second Ricci tensor. Notice that only the second Ricci tensor is symmetric in its indices: $\tilde R_{\mu\nu}=\tilde R_{\nu\mu}$. There are two more contractions of the generalized curvature tensor: $R^\lambda_{\;\;\lambda\mu\nu}$ and $R^{\;\;\;\;\lambda}_{\mu\nu\;\;\lambda}$. However, the latter one identically vanishes while the former one is a linear combination of the contractions $R_{\mu\nu}$ and $\tilde R_{\mu\nu}$:

\bea R^\lambda_{\;\;\lambda\mu\nu}=2\left(R_{\mu\nu}-\tilde R_{\mu\nu}\right)=2\left(\hat\nabla_\mu Q_\nu-\hat\nabla_\nu Q_\mu\right).\nonumber\eea

From equations \eqref{1-ricci-t} and \eqref{2-ricci-t} it follows that,

\bea R_{(\mu\nu)}=\tilde R_{\mu\nu},\;R_{[\mu\nu]}=\hat\nabla_\mu Q_\nu-\hat\nabla_\nu Q_\mu.\nonumber\eea Besides, for the Einstein's tensor $G_{\mu\nu}:=R_{\mu\nu}-g_{\mu\nu}R/2$ we obtain that, $G_{(\mu\nu)}=\tilde G_{\mu\nu}\equiv\tilde R_{\mu\nu}-g_{\mu\nu}R/2.$



\subsection{Weyl gauge symmetry}
 

Weyl gauge symmetry (WGS) or invariance under local changes of scale, is a manifest symmetry of $\tilde W_4$ spaces. The geometric laws that define $\tilde W_4$, among which is the nonmetricity condition \eqref{vect-nm}, are invariant under Weyl gauge transformations \eqref{gauge-t}, which represent a particular case of \eqref{gen-gauge-t}. Under \eqref{gauge-t}:

\bea &&\{^\alpha_{\mu\nu}\}\rightarrow\{^\alpha_{\mu\nu}\}+\left(\delta^\alpha_\mu\der_\nu+\delta^\alpha_\nu\der_\mu-g_{\mu\nu}\der^\alpha\right)\ln\Omega,\nonumber\\
&&L^\alpha_{\;\;\mu\nu}\rightarrow L^\alpha_{\;\;\mu\nu}-\left(\delta^\alpha_\mu\der_\nu+\delta^\alpha_\nu\der_\mu-g_{\mu\nu}\der^\alpha\right)\ln\Omega,\nonumber\eea so that the generalized affine connection \eqref{gen-aff-c} is unchanged by the Weyl rescalings: $\Gamma^\alpha_{\;\;\mu\nu}\rightarrow\Gamma^\alpha_{\;\;\mu\nu}.$ This means that the generalized curvature tensor $R^\alpha_{\;\;\sigma\mu\nu}$ in \eqref{gen-curv-t} and the generalized Ricci tensor, $R_{\mu\nu}\equiv R^\lambda_{\;\;\mu\lambda\nu}$, are unchanged as well, $R^\alpha_{\;\;\mu\sigma\nu}\rightarrow R^\alpha_{\;\;\mu\sigma\nu}$, $R_{\mu\nu}\rightarrow R_{\mu\nu}$, while the generalized curvature scalar transforms as, $R\rightarrow\Omega^{-2}R.$ It can be straightforwardly demonstrated, that the Bianchi identities are gauge invariant expressions as well.

Another important quantity is the nonmetricity field strength. It is a traceless second-rank tensor with coordinate components,

\bea Q_{\mu\nu}:=2\nabla_{[\mu}Q_{\nu]}=\nabla_\mu Q_\nu-\nabla_\nu Q_\mu=\der_\mu Q_\nu-\der_\nu Q_\mu,\label{qmn}\eea which under the gauge transformations \eqref{gauge-t} it is not transformed. The quantity \eqref{qmn} represents that part of the curvature which is due to nonmetricity of $\tilde W_4$ space.



\section{Gauge invariant theory of gravity}\label{sect-ginv}


In this paper we shall consider the class of gauge invariant theories of gravity that are given by the gravitational action \eqref{g-action'}. Let us further modify this action through substituting the curvature scalar from \eqref{gen-curv-sc} and by explicitly writing the gauge derivative: 

\bea (\der^*\phi)^2=(\der\phi)^2-\phi\der_\mu\phi Q^\mu+\frac{\phi^2}{4}Q_\mu Q^\mu.\label{explicit-der}\eea The action \eqref{g-action'} then reads,

\bea &&S_g=\frac{1}{2}\int d^4x\sqrt{-g}\left[\phi^2\hat R+\omega(\der\phi)^2+\frac{\omega-6}{4}\phi^2Q_\mu Q^\mu\right.\nonumber\\
&&\;\;\;\;\;\;\;\;\;\;\;\;\;\;\;\;\;\;\;\;\;\;\left.+\frac{\omega-6}{2}\phi^2\hat\nabla_\mu Q^\mu-\frac{\beta^2}{2}Q^2-\frac{\lambda}{4}\phi^4\right].\label{fin-action}\eea Variation of the above action with respect to the metric leads to the following EOM:

\bea &&{\cal E}_{\mu\nu}:=\hat G_{\mu\nu}-\frac{1}{\phi^2}\left(\hat\nabla_\mu\hat\nabla_\nu-g_{\mu\nu}\hat\nabla^2\right)\phi^2\nonumber\\
&&\;\;\;\;\;+\frac{\omega}{\phi^2}\left[\der_\mu\phi\der_\nu\phi-\frac{1}{2}g_{\mu\nu}(\der\phi)^2\right]\nonumber\\
&&\;\;\;\;\;+\frac{\omega-6}{4}\left(Q_\mu Q_\nu-\frac{1}{2}g_{\mu\nu}Q_\lambda Q^\lambda\right)\nonumber\\
&&\;\;\;\;\;-\frac{\omega-6}{2\phi^2}\left[\der_{(\mu}\phi^2Q_{\nu)}-\frac{1}{2}g_{\mu\nu}\der_\lambda\phi^2Q^\lambda\right]\nonumber\\
&&\;\;\;\;\;-\frac{\beta^2}{\phi^2}\left(Q^{\;\;\lambda}_\mu Q_{\nu\lambda}-\frac{1}{4}g_{\mu\nu}Q^2\right)=-\frac{\lambda}{8}\phi^2g_{\mu\nu},\label{einst-eom}\eea where, during the variation procedure we took into account the following useful expression:

\bea &&\delta_{\bf g}\left(\hat\nabla_\lambda Q^\lambda\right)=\hat\nabla_{(\mu}Q_{\nu)}\delta g^{\mu\nu}+Q_{(\mu}\hat\nabla_{\nu)}\delta g^{\mu\nu}\nonumber\\
&&\;\;\;\;\;\;\;\;\;\;\;\;\;\;\;\;\;\;\;\;\;\;-\frac{1}{2}g_{\mu\nu}Q^\lambda\hat\nabla_\lambda\delta g^{\mu\nu},\label{usef-var}\eea where $\delta_{\bf g}$ means variation with respect to the metric. Variation of \eqref{fin-action} with respect to $Q_\mu$ leads to the following inhomogeneous Proca EOM:\footnote{Equation \eqref{max-eom} can be rewritten in the following fully equivalent form: $$\hat\nabla^\nu Q_{\mu\nu}+m^2_QQ_\mu=j^\text{eff}_\mu,$$ where we introduced the point-dependent square mass of the Proca field $m^2_Q$ and an effective current $j^\text{eff}_\mu$: $$m^2_Q\equiv\frac{6-\omega}{4\beta^2}\phi^2,\;j^\text{eff}_\mu\equiv\frac{6-\omega}{4\beta^2}\der_\mu\phi^2,$$ respectively.\label{foot}}

\bea \hat\nabla^\mu Q_{\mu\nu}=\frac{6-\omega}{4\beta^2}\left(\phi^2Q_\nu-\der_\nu\phi^2\right),\label{max-eom}\eea meanwhile, variation with respect to $\phi$ yields:

\bea &&\hat R+\omega\frac{(\der\phi)^2}{\phi^2}-\frac{\omega}{2}\frac{\hat\nabla^2\phi^2}{\phi^2}\nonumber\\
&&+\frac{\omega-6}{4}Q_\mu Q^\mu+\frac{\omega-6}{2}\hat\nabla_\mu Q^\mu-\frac{\lambda}{2}\phi^2=0.\label{phi-eom}\eea  

If take the LC divergence of Eq. \eqref{max-eom}, recalling that $Q_{\mu\nu}=-Q_{\nu\mu}$ $\Rightarrow\hat\nabla^\mu\hat\nabla^\mu Q_{\mu\nu}=0$, we get that

\bea \frac{6-\omega}{4\beta^2}\left[\hat\nabla_\mu\left(\phi^2Q^\mu\right)-\hat\nabla^2\phi^2\right]=0.\label{kg-eom}\eea This equation is obtained as well if compare \eqref{phi-eom} with the trace of \eqref{einst-eom}.



\subsection{Particular cases found in the bibliography}


The class of gauge invariant theories given by the action \eqref{fin-action} includes the Weyl geometric theoretical frameworks developed in \cite{dirac-1973} where a formalism based in the revival of Weyl geometry is proposed in order to explain the possible variation of the gravitational constant with time, and in \cite{smolin-1979}, which corresponds to a particular case of \eqref{fin-action} when $\omega=1/c$. In this approach a gauge invariant extension of GR based on Weyl geometry is proposed to look for short-distance effects of gravity.\footnote{In \cite{smolin-1979} the quadratic terms $R_{\mu\nu}R^{\mu\nu}$ and $R^2$ are dropped as they induce unphysical poles in the graviton propagator and, besides, they do not contribute to the low-energy phenomenology.} The formalism investigated in the Ref. \cite{ghilen-1} (see the related references \cite{harko-prd-2023, ghilen-2023, ghilen-2, ghilen-3, ghilen-4}) also belongs in the class of gauge invariant theories \eqref{fin-action} if set $\omega=0$. In this case the possible modifications of the SM by replacing Riemann by Weyl geometry are investigated. The approach in Ref. \cite{bars-2014} (see also Ref. \cite{bars-2014-1}), corresponds to the particular case of \eqref{g-action'} with $\omega=6$. The formalism of Ref. \cite{cheng-1988, cheng-arxiv}, is contained in the above class as well, if in \eqref{g-action'} replace the scalar field $\phi$ by the multicomponent (complex) scalar field $\vphi$, such that $\phi^2\rightarrow|\vphi|^2=\vphi^\dag\vphi$, $(\der^*\phi)^2\rightarrow|\der^*\vphi|^2$, etc. In this approach a gauge invariant field theory for EW and gravitational interactions in Weyl background space is explored.



\section{Matter coupling}\label{sect-lemma-1}


Let us consider a matter piece of action,

\bea S_m=\int d^4x\sqrt{-g}{\cal L}_m(\psi,\nabla^*\psi,g),\label{m-action}\eea where $\psi$ denotes any minimally coupled matter fields. Variation of the above action with respect to the metric leads to,

\bea \delta_{\bf g}S_m=-\frac{1}{2}\int d^4x\sqrt{-g}\delta g^{\mu\nu}T^{(m)}_{\mu\nu},\label{var-action-m-1}\eea where $T^{(m)}_{\mu\nu}:=-(2/\sqrt{-g})\der(\sqrt{-g}{\cal L}_m)/\der g^{\mu\nu},$ is the stress-energy tensor of the matter fields. The equations of motion which follow by varying the overall action $S_\text{tot}=S_g+S_m$ with respect to the fields $g_{\mu\nu}$, $Q_\mu$ and $\phi$, read:

\bea {\cal E}_{\mu\nu}=\frac{1}{\phi^2}T^{(m)}_{\mu\nu}-\frac{\lambda}{8}\phi^2g_{\mu\nu},\label{einst-mat-eom}\eea where ${\cal E}_{\mu\nu}$ is defined in Eq. \eqref{einst-eom}, plus the EOM \eqref{max-eom} and \eqref{phi-eom} which, thanks to the minimal coupling of the matter fields, are not modified. Meanwhile, if compare the trace of \eqref{einst-mat-eom} with equation \eqref{phi-eom}, one gets

\bea \hat\nabla_\mu\left(\phi^2Q^\mu\right)-\hat\nabla^2\phi^2=\frac{2}{\omega-6}\,T^{(m)},\label{kg-mat-eom}\eea which replaces \eqref{kg-eom}. In this equation $T^{(m)}=g^{\mu\nu}T^{(m)}_{\mu\nu}$ is the trace of the matter SET. 


The following lemma takes place:

\begin{lemma} Let the action of a class of gauge invariant theories of gravity over background space $\tilde W_4$ with nonmetricity $Q_\mu$, be given by the action

\bea S_\text{tot}=S_g+S_m,\label{weyl-tot-action}\eea where $S_g$ is defined by \eqref{fin-action} while $S_m$ is defined by \eqref{m-action}, so that the derived EOM are \eqref{max-eom}, \eqref{phi-eom} and \eqref{einst-mat-eom}. Then, (i) only matter fields with traceless SET couple to gravity and (ii) these follow geodesics of Riemann geometry.\label{lemma-1}\end{lemma}

{\bf Proof.} The proof of this lemma is in two parts. First we shall proof that only matter fields with vanishing SET trace satisfy the EOM \eqref{max-eom}, \eqref{phi-eom} and \eqref{einst-mat-eom}. Then we shall proof that these matter fields follow null geodesics of Riemann space $V_4$.


Let us demonstrate that only matter fields with traceless SET: $T^{(m)}=0$, satisfy the EOM \eqref{max-eom}, \eqref{phi-eom} and \eqref{einst-mat-eom}. Given that equations \eqref{max-eom} and \eqref{phi-eom} are not modified by the presence of matter, and that the divergence of the left-hand side (LHS) of Eq. \eqref{max-eom} identically vanishes since the nonmetricity field strength $Q_{\mu\nu}$ is antisymmetric, then the divergence of its right-hand side (RHS) vanishes as well, as shown in Eq. \eqref{kg-eom}. If compare equations \eqref{kg-eom} and \eqref{kg-mat-eom} it follows that $T^{(m)}=0$. This means that only matter with traceless SET: radiation and massless fields, couple to gravity in this class of theory. Fields with $T^{(m)}\neq 0$ do not obey the EOM.


Let us now demonstrate that radiation and massless matter fields follow null geodesics of Riemann geometry. Here, for sake of simplicity, we use the Brans-Dicke (BD) notation so that, in equations \eqref{max-eom}, \eqref{phi-eom}, \eqref{kg-eom} and \eqref{einst-mat-eom}, we make the following replacements: $\phi^2\rightarrow\vphi$ and $\omega\rightarrow4\bar\omega$.\footnote{In order to get the correct sign of the BD coupling constant we have to set $\bar\omega=-\omega_\text{BD}$.} Besides, we decompose the tensor ${\cal E}_{\mu\nu}$ defined in \eqref{einst-eom} in the following form:

\bea {\cal E}_{\mu\nu}=\frac{1}{\vphi}\left(\hat{\cal G}_{\mu\nu}+{\cal Q}_{\mu\nu}+{\cal F}_{\mu\nu}\right),\label{a-rewrite}\eea where

\bea &&\hat{\cal G}_{\mu\nu}\equiv\vphi\hat G_{\mu\nu}-\left(\hat\nabla_\mu\hat\nabla_\nu-g_{\mu\nu}\hat\nabla^2\right)\vphi\nonumber\\
&&\;\;\;\;\;\;\;\;\;\;\;+\frac{\bar\omega}{\vphi}\left[\der_\mu\vphi\der_\nu\vphi-\frac{1}{2}g_{\mu\nu}\left(\der\vphi\right)^2\right],\nonumber\\
&&{\cal Q}_{\mu\nu}\equiv\left(\bar\omega-\frac{3}{2}\right)\left[\vphi\left(Q_\mu Q_\nu-\frac{1}{2}g_{\mu\nu}Q_\lambda Q^\lambda\right)\right.\nonumber\\
&&\;\;\;\;\;\;\;\;\;\;\;\left.-\der_\mu\vphi Q_\nu-\der_\nu\vphi Q_\mu+g_{\mu\nu}\der_\lambda\vphi Q^\lambda\right],\nonumber\\
&&{\cal F}_{\mu\nu}\equiv-\beta^2\left(Q^{\;\;\lambda}_{\mu}Q_{\nu\lambda}-\frac{1}{4}g_{\mu\nu}Q_{\lambda\sigma}Q^{\lambda\sigma}\right).\label{a-def}\eea It is not difficult to show that,

\bea &&\hat\nabla^\mu\hat{\cal G}_{\mu\nu}=\hat\nabla_\nu\vphi\left[\bar\omega\frac{\hat\nabla^2\vphi}{\vphi}-\frac{\bar\omega}{2}\frac{(\der\vphi)^2}{\vphi^2}-\frac{1}{2}\hat R\right]\nonumber\\
&&=\hat\nabla_\nu\vphi\left\{\left(\bar\omega-\frac{3}{2}\right)\left[\hat\nabla_\mu Q^\mu+\frac{1}{2}Q_\mu Q^\mu\right]-\frac{\lambda}{4}\vphi\right\},\label{a-div-g}\eea where in order to go from first to second lines we used the EOM \eqref{phi-eom}. Besides, above we took into account the Bianchi identity in the form, $\hat\nabla^\mu\hat G_{\mu\nu}=0$, and the following useful expression:

\bea \left(\hat\nabla^2\hat\nabla_\mu-\hat\nabla_\mu\hat\nabla^2\right)\vphi=\hat R_{\mu\nu}\hat\nabla^\nu\vphi.\label{a-usef-1}\eea After some algebra it can be shown that,

\bea &&\hat\nabla^\mu{\cal Q}_{\mu\nu}=-\left(\bar\omega-\frac{3}{2}\right)\left[\hat\nabla_\nu\vphi\left(\hat\nabla_\mu Q^\mu+\frac{1}{2}Q_\mu Q^\mu\right)\right.\nonumber\\
&&\;\;\;\;\;\;\;\;\;\;\;\;\;\;\;\;\;\left.+J^\mu Q_{\mu\nu}+\left(\hat\nabla^\mu J_\mu\right)Q_\nu\right]\nonumber\\
&&\;\;\;\;\;\;\;\;\;\;\;\;=-\hat\nabla_\nu\vphi\left(\bar\omega-\frac{3}{2}\right)\left[\hat\nabla_\mu Q^\mu+\frac{1}{2}Q_\mu Q^\mu\right]\nonumber\\
&&\;\;\;\;\;\;\;\;\;\;\;\;\;\;\;\;-\left(\bar\omega-\frac{3}{2}\right)J^\mu Q_{\mu\nu},\label{a-div-q}\eea where we have introduced the shorthand notation $J_\mu\equiv\der_\mu\vphi-\vphi Q_\mu$, and in the last step we took into account the EOM \eqref{max-eom}: $\beta^2\hat\nabla^\mu Q_{\mu\nu}=(\bar\omega-3/2)J_\nu$, so that $\hat\nabla^\mu J_\mu\propto\hat\nabla^\mu\hat\nabla^\nu Q_{\nu\mu}=0$. We have that,

\bea &&\hat\nabla^\mu{\cal F}_{\mu\nu}=\left(\bar\omega-\frac{3}{2}\right)J^\mu Q_{\mu\nu}-\beta^2\left[2Q^{\mu\lambda}\hat\nabla_{[\mu}\hat\nabla_{\nu]}Q_\lambda\right.\nonumber\\
&&\;\;\;\;\;\;\;\;\;\;\;\;\;\;\left.-Q^{\mu\lambda}\hat\nabla_{[\mu}\hat\nabla_{\lambda]}Q_\nu\right].\label{a-div-f}\eea Now, if took into account the following useful expression:

\bea 2\hat\nabla_{[\mu}\hat\nabla_{\nu]}Q_\lambda=-\hat R^\sigma_{\;\;\lambda\mu\nu}Q_\sigma,\label{a-usef-2}\eea and the definition of the Riemann-Christoffel curvature tensor of $V_4$ space \eqref{lc-curv-t}, the following equation takes place:

\bea Q^{\mu\lambda}\hat R^\sigma_{\;\;\lambda\mu\nu}=\frac{1}{2}Q^{\mu\lambda}\hat R^\sigma_{\;\;\nu\mu\lambda},\nonumber\eea so that the expression within square brackets in \eqref{a-div-f} vanishes:

\bea 2Q^{\mu\lambda}\hat\nabla_{[\mu}\hat\nabla_{\nu]}Q_\lambda-Q^{\mu\lambda}\hat\nabla_{[\mu}\hat\nabla_{\lambda]}Q_\nu=0,\nonumber\eea which leads to

\bea \hat\nabla^\mu{\cal F}_{\mu\nu}=\left(\bar\omega-\frac{3}{2}\right)J^\mu Q_{\mu\nu}.\label{a-div-f'}\eea Taking into account equations \eqref{a-rewrite}, \eqref{a-def}, \eqref{a-div-g}, \eqref{a-div-q} and \eqref{a-div-f'}, we finally obtain:

\bea \hat\nabla^\mu\left(\vphi{\cal E}_{\mu\nu}\right)=\hat\nabla^\mu\hat{\cal G}_{\mu\nu}+\hat\nabla^\mu{\cal Q}_{\mu\nu}+\hat\nabla^\mu{\cal F}_{\mu\nu}=-\frac{\lambda}{4}\vphi\hat\nabla_\nu\vphi.\label{qed}\eea Hence, the following vanishing divergence takes place:

\bea \hat\nabla^\mu\left(\phi^2{\cal E}_{\mu\nu}+\frac{\lambda}{8}\phi^4g_{\mu\nu}\right)=0.\label{div-0}\eea If further consider the gravitational EOM \eqref{einst-mat-eom}, the above equation entails that the standard GR conservation equation,\footnote{Equation \eqref{gr-cons-eq} is consistent with the well-known result that the Riemannian null geodesic equations are invariant under the Weyl rescalings \eqref{gauge-t}, as shown in appendix D of Ref. \cite{wald-book}.}

\bea \hat\nabla^\mu T^{(m)}_{\mu\nu}=0,\label{gr-cons-eq}\eea is satisfied. Since, as demonstrated in the first part of the proof, only matter fields with traceless SET obey the EOM \eqref{einst-mat-eom}, \eqref{max-eom} and \eqref{phi-eom}, then Eq. \eqref{gr-cons-eq} means that massless matter fields respond only to the curvature of Riemann space $V_4$, i. e., that these follow null geodesics of Riemann geometry. {\bf Q.E.D.} 

The following corollary of lemma \ref{lemma-1} takes place:

\begin{corollary} Massless fields -- the only matter fields which satisfy the EOM of the class of gauge invariant gravitational theories given by the gravitational action \eqref{fin-action} -- do not interact with the nonmetricity vector $Q_\mu$.\label{coro-l1}\end{corollary} 

The physical consequences of this corollary discourage the potential influence of vectorial nonmetricity and of gauge symmetry on the gravitational phenomena. Actually, the fact that the only matter fields that can be included in the class of theories \eqref{g-action'}: radiation and massless fields, do not interact with the nonmetricity, means that $Q_\mu$ has not effective impact on these fields and may be ignored.



\section{Low-energy phenomenology}\label{sect-low-energy}


Due to fulfillment of lemma \ref{lemma-1} (and of its corollary,) we must replace the arbitrary matter Lagrangian ${\cal L}_{m}$ by the radiation Lagrangian ${\cal L}_\text{rad}$, which leads to the following traceless SET:

\bea T^\text{rad}_{\mu\nu}=-\frac{2\der(\sqrt{-g}{\cal L}_\text{rad})}{\sqrt{-g}\der g^{\mu\nu}}=\frac{4}{3}\rho_\text{rad}\left(u_\mu u_\nu+\frac{1}{4}g_{\mu\nu}\right),\label{rad-set}\eea where $u^\mu=\delta^\mu_0$ is the fourth-velocity of co-moving observers. The action piece $S_\text{rad}=\int d^4x\sqrt{-g}{\cal L}_\text{rad}$ contains contributions from all of the SM fields prior to EW symmetry breaking, i. e., when these are massless fields. Besides, to the action \eqref{weyl-tot-action} we need to add the $SU(2)\times U(1)$ symmetry breaking piece $S_H=\int d^4x\sqrt{-g}{\cal L}_H$, where the following gauge invariant Higgs Lagrangian is assumed:

\bea {\cal L}_H=\frac{1}{2}|D^*_gH|^2+\frac{\lambda'}{2}\left[|H|^2-v^2(x)\right]^2,\label{weyl-higgs-lag}\eea where $v(x)=v_0\phi(x)$ is the point-dependent mass parameter ($\lambda'$ and $v_0$ are dimensionless constants.) In \eqref{weyl-higgs-lag} we adopted the following notation: $|D^*_gH|^2\equiv g^{\mu\nu}(D^{g*}_\mu H)^\dag D^{g*}_\nu H$, where $D^{g*}_\mu=D^g_\mu-Q_\mu/2$ and $D^g_\mu$ is defined in \eqref{ew-gauge-der}. 

In what follows we shall consider the following gauge invariant action:

\bea S_\text{tot}=S_g+S_H+S_\text{rad},\label{tot-action'}\eea where the matter action for radiation $S_\text{rad}$ and the EW symmetry breaking action $S_H$ have been defined in the text above Eq. \eqref{weyl-higgs-lag}, while the gravitational action $S_g=\int d^4x\sqrt{-g}{\cal L}_g$ is given by action \eqref{fin-action}, whose associated Lagrangian ${\cal L}_g$ can be rewritten as

\bea {\cal L}_g=\frac{1}{2}\left[\phi^2\hat R+6(\der\phi)^2-\frac{\lambda}{4}\phi^4\right]+{\cal L}_S,\label{g-lag}\eea where

\bea {\cal L}_S=\frac{\beta^2}{4}\left[-Q^2+\frac{\omega-6}{2\beta^2}\phi^2\left(Q_\mu-\frac{\der_\mu\phi^2}{\phi^2}\right)^2\right],\label{stueck-grav-lag}\eea stands for the Stueckelberg-type Lagrangian\footnote{Notice that ${\cal L}_S$ differs from the standard Stueckelberg Lagrangian in the absence of a gauge fixing term \cite{stueck-1, stueck-2, stueck-3, stueck-4}. Yet it is not a typical Proca Lagrangian thanks to the gradient $\der_\mu\phi^2/\phi^2$ within round brackets squared. This leads to the Lagrangian density $\sqrt{-g}{\cal L}_S$ being gauge invariant in contrast to just Proca term which is not gauge invariant.} of the Proca field $Q_\mu$ and we used the notation $(a_\mu+b_\mu)^2\equiv(a_\mu+b_\mu)(a^\mu+b^\mu)$. 

The independent gravitational EOM that can be derived from \eqref{tot-action'} are the Einstein's EOM:\footnote{For simplicity we omit the Higgs field.}

\bea &&{\cal E}_{\mu\nu}=\frac{1}{\phi^2}T^\text{rad}_{\mu\nu}-\frac{\lambda}{8}\phi^2g_{\mu\nu},\label{S-einst-eom}\eea which is obtained by varying the action \eqref{tot-action'} with respect to the metric, and the inhomogeneous Proca equation \eqref{max-eom} (see footnote \ref{foot},) which is obtained by varying with respect to the nonmetricity vector $Q_\mu$. 

The scalar field EOM \eqref{phi-eom}, which is obtained by varying the action \eqref{tot-action'} with respect to $\phi$, is not an independent equation since it can be obtained by substituting the vanishing LC covariant divergence of \eqref{max-eom}: $\hat\nabla^\nu(\hat\nabla^\mu Q_{\mu\nu})=0$ $\Rightarrow\hat\nabla^\mu(\phi^2Q_\mu)-\hat\nabla^2\phi^2=0$, into the trace of Einstein's equation \eqref{S-einst-eom}, recalling that $T^\text{rad}=0$. This means that the scalar field does not satisfy any specific EOM; i. e., $\phi$ can be chosen at will. Hence, $\phi$ is not a dynamical DOF and the coupling constant $\omega$ does not affect the measured Newton's constant. Besides, given that the nonmetricity vector $Q_\mu$ does not interact with the matter fields, it does not modify the measured gravitational constant either. For these reasons the measured Newton's constant in the class of theories \eqref{tot-action'} corresponds to the tensor gravitational force. It is given by,

\bea 8\pi G_N(x)=\frac{1}{\phi^2(x)},\label{newton-c}\eea so that it depends on spacetime point like in the BD theory.\footnote{In contrast, in the BD theory, since the scalar field $\phi$ is a dynamical (gravitational) degree of freedom, the measured gravitational constant is indeed modified by $\phi$ through the BD coupling constant $\omega_\text{BD}$ \cite{dicke-1962, bd-theory, fujii-book}: $$8\pi G_N=\frac{1}{\phi^2}\left(\frac{4+2\omega_\text{BD}}{3+2\omega_\text{BD}}\right).$$}

In order to understand why the nonmetricity vector field $Q_\mu$, being a (non-Riemannian) part of the gravitational field, does not interact with radiation: the only matter degrees of freedom allowed by the theory \eqref{g-lag}, let us bring into attention the gravitational spectrum of this gauge invariant theory. Because the scalar degree of freedom associated with $\phi$ is not dynamical, the gravitational spectrum of \eqref{g-lag} consists of two degrees of freedom of the massless graviton plus three degrees of freedom of the massive field $Q_\mu$, whose effective mass squared is given by (see footnote \ref{foot},)

\bea m^2_Q(x)=\frac{6-\omega}{4\beta^2}\phi^2(x)=\frac{6-\omega}{4\beta^2}M^2_\text{pl}(x).\label{eff-mass}\eea In this equation $M_\text{pl}(x)=1/8\pi G_N(x)$ stands for the point-dependent effective Planck mass. This quantity sets the grand unification scale point by point in spacetime. Hence, unless either $\omega=6$ or $\beta^2\rightarrow\infty$, the effective mass $m_Q(x)\sim M_\text{pl}(x)$, meaning that the nonmetricity field is decoupled from the low-energy gravitational spectrum. In this case, thanks to the fact that the Stueckelberg-type Lagrangian density $\sqrt{-g}{\cal L}_S$ \eqref{stueck-grav-lag} is gauge invariant itself, we may dispense with the Lagrangian ${\cal L}_S$ without affecting the gauge symmetry of the resulting gravitational Lagrangian density $\sqrt{-g}{\cal L}_g$ in \eqref{g-lag}. In the opposite end stands the case when $\omega=6$, which corresponds to massless $Q_\mu$. In this case the nonmetricity field amounts to an additional radiation (matter) field propagating in the background Riemann space $V_4$, so that it may be disregarded as well. Hence, the low-energy gravitational spectrum of the gauge invariant theory \eqref{g-lag} is the same as in GR: It consists of the two polarizations of the graviton exclusively.

In this paper we are interested in the low-energy phenomenology so that we ignore the Stueckelberg-type Lagrangian ${\cal L}_S$ in \eqref{stueck-grav-lag}. The resulting gauge invariant (effective) gravitational Lagrangian reads,

\bea {\cal L}^\text{eff}_g=\frac{1}{2}\left[\phi^2\hat R+6(\der\phi)^2-\frac{\lambda}{4}\phi^4\right].\label{eff-lag}\eea It coincides with the particular case when in \eqref{fin-action} we set $\omega=6$, $\beta^2=0$. This is the well-known Lagrangian of a conformally coupled scalar. If we want to go beyond the low-energy phenomenology, consideration of higher curvature terms is mandatory. Nevertheless, in such a case lemma \ref{lemma-1} and its corollary are not satisfied in general.\footnote{See, however, the related discussion in section \ref{sect-discu}.}

From the overall action

\bea S_\text{tot}=\int d^4x\sqrt{-g}{\cal L}^\text{eff}_g+S_\text{rad},\label{tot-action}\eea with ${\cal L}^\text{eff}_g$ given by \eqref{eff-lag}, the following EOM are derived:

\bea &&\hat G_{\mu\nu}-\frac{1}{\phi^2}\left(\hat\nabla_\mu\hat\nabla_\nu-g_{\mu\nu}\hat\nabla^2\right)\phi^2\nonumber\\
&&+\frac{6}{\phi^2}\left[\der_\mu\phi\der_\nu\phi-\frac{1}{2}g_{\mu\nu}(\der\phi)^2\right]=\frac{1}{\phi^2}T^\text{rad}_{\mu\nu}-\frac{\lambda}{8}\phi^2g_{\mu\nu},\nonumber\\
&&\hat R+6\frac{(\der\phi)^2}{\phi^2}-3\frac{\hat\nabla^2\phi^2}{\phi^2}-\frac{\lambda}{2}\phi^2=0,\label{eff-eom}\eea together with the conservation equation $\hat\nabla^\mu T^\text{rad}_{\mu\nu}=0.$ Since the trace of the radiation SET vanishes ($T^\text{rad}=0$,) the second EOM above -- which is derived by varying \eqref{tot-action} with respect to $\phi$ -- is not an independent equation since it coincides with the trace of the Einstein's equation in \eqref{eff-eom}. Hence, the scalar field does not satisfy an independent EOM. 

In general $\phi$ can be set equal to any nonvanishing (continuous) function $\phi=\phi(t,\vec{x})$ or to any constant $\phi=\phi_0$ without conflict with the EOM \eqref{eff-eom}. This means the $\phi$ is a non-dynamical field.



\section{Gauge freedom: The many-worlds interpretation}\label{sect-g-freedom}


We have argued that the usual EM-inspired interpretation of gauge symmetry according to which a specific gauge choice carries no physical consequences as the different gauges describe the same physical state, is not appropriate in the case of gauge invariant theories of gravity. In this case the conformal transformations of the metric affect the measuring scales and, hence, the way we do measurements of time and length. Besides, these affect the scalar density of fermions and related quantities as well. We need a different perspective on gauge invariance and on what gauge fixing means in this case. 
 
Here we shall develop an alternative understanding of gauge symmetry and of gauge fixing in gravitational theories which, despite obvious differences, bears resemblance with the many-worlds interpretation of quantum physics \cite{everett, dewitt, kent, barvinsky, omnes, tegmark, garriga, zurek, tegmark-nature, page}. Since the scalar field $\phi$ can be any smooth function, choosing a specific gauge, labeled ``$j$,'' means choosing a specific function $\phi_j$ ($j\in\mathbb{N}$). Let us represent a given gauge by,

\bea {\cal G}_j:\left\{{\cal M}_4\in V_4,\,g_{\mu\nu},\,\phi_j\,|\,{\cal S}_j,\,{\cal C},...\right\},\label{j-gauge}\eea where, due to lemma \ref{lemma-1} and its corollary (see also the above discussion on the low-energy phenomenology,) we have replaced the starting Weyl geometric spacetime structure by the effective Riemannian spacetime structure: $\tilde W_4\rightarrow V_4$. In \eqref{j-gauge} ${\cal S}_j$ represents the set of relevant measured point-dependent ``constants'' of nature $${\cal S}_j=\left\{M^2_{\text{pl},j}(x),\,v_j(x),\,\Lambda_j(x)\right\},$$ where the effective (point-dependent) Planck mass reads, $$M^2_{\text{pl},j}(x)=\frac{1}{8\pi G_{N,j}(x)}=\phi^2_j(x),$$ while the point-dependent mass parameter in the gauge invariant Higgs Lagrangian \eqref{weyl-higgs-lag} and the effective cosmological constant are given by, $$v_j(x)=v_0\phi_j(x),\,\Lambda_j(x)=\frac{\lambda}{8}\phi^2_j(x),$$ respectively. In the above definition of ${\cal G}_j$ we have included the set of measured gauge invariant constants of nature: ${\cal C}=\{\hbar,\,c,\,e,...\},$ where $\hbar$ is the Plack constant, $c$ is the speed of light, $e$ is the EM charge of the electron and the ellipsis stand for other physical constants which are not transformed by the gauge transformations \eqref{gauge-t}. The ellipsis in \eqref{j-gauge} represent other possible relevant measured quantities in the theory under consideration.

Each gauge carries a potential description or representation of the world. Although the gravitational laws \eqref{eff-eom} are gauge invariant, once a gauge is picked out these laws lose the manifest gauge symmetry. Means that the laws look different in different ${\cal G}_j$-s: Since $\phi_j(x)$ is different in different gauges, the measured Newton's constant (inverse of the point-dependent Planck mass squared,) the Higgs mass parameter and the energy density of vacuum $\rho^\text{vac}_j(x)=\Lambda_j(x)M^2_{\text{pl},j}(x)$, among other measured quantities, are different in the different gauges. Yet a residual gauge symmetry remains in the following sense: Any gauge ${\cal G}_i$ is related with any other ${\cal G}_j$ through gauge transformations,

\bea g_{\mu\nu}\rightarrow\Omega^2g_{\mu\nu},\,\phi_i\rightarrow\Omega^{-1}\phi_j,\label{gauge-t'}\eea plus appropriate transformations of the remaining fields. By means of \eqref{gauge-t'} a given gauge transforms into another gauge: ${\cal G}_i\rightarrow{\cal G}_j$, while ${\cal G}_j\rightarrow{\cal G}_i$ through the inverse transformations. 

The overall picture consists of a conformal equivalence class of gauges which is generated by the infinity of possible choices of the scalar $\phi_j$:

\bea {\cal K}=\left\{{\cal G}_1,{\cal G}_2,...,{\cal G}_j,...,{\cal G}_N|\;j\in\mathbb{N}\right\},\label{c-class}\eea where the general element of the class ${\cal G}_j$ is given by \eqref{j-gauge} and $N\rightarrow\infty$. Any two elements of the conformal equivalence class ${\cal K}$ \eqref{c-class} are linked by gauge transformations \eqref{gauge-t'}. Imagine a number $N\rightarrow\infty$ of identical copies ${\cal W}_j$ of our world. Now let us associate with each copy a physical/geometrical description given by an element of ${\cal K}$: ${\cal G}_j\leftrightarrow{\cal W}_j$. We end up with $N$ different worlds which have been subject to different descriptions. Without loss of generality -- putting aside obvious philosophical counter-arguments -- we may establish an equivalence between given worlds and their physical/geometrical descriptions: ${\cal W}_j\Leftrightarrow{\cal G}_j$. Hence, the conformal equivalence class ${\cal K}$ is equivalent to the class of potential worlds: $\left\{{\cal W}_1,{\cal W}_2,...,{\cal W}_j,...,{\cal W}_N|\;j\in\mathbb{N}\right\}.$ It is in this sense that we establish a parallel between our interpretation of gauge fixing and the many-worlds picture.

This classic gravitational version of the many-worlds interpretation of quantum physics is interesting because it provides a different perspective on the relation between theory and experiment. Usually experiment is useful in order to corroborate the theoretical predictions made on the basis of given theoretical framework. According to the present approach experiment allows one to determine which one of the infinitely many gauges is the one which better describes our universe, through associating experimental values to the measured quantities, in particular to the (point-dependent) constants of nature.



\subsection{General relativity gauge}\label{subsect-gr-gauge}


Among the infinity of possible gauges there is one which is singular. If in the action \eqref{tot-action'} make the following choice of the scalar field: $\phi=M_\text{pl}$, where $M_\text{pl}$ is the Planck mass, one obtains\footnote{The EH action \eqref{gr-action} can be obtained from \eqref{tot-action'} through the gauge transformations \eqref{gauge-t'}: $g_{\mu\nu}\rightarrow\Omega^2g_{\mu\nu}$, $\phi\rightarrow\Omega^{-1}M_\text{pl}.$ The inverse transformations map the GR action \eqref{gr-action} back into \eqref{tot-action'}. }

\bea S_\text{tot}=\frac{1}{2}\int d^4x\sqrt{-g}\left[M^2_\text{pl}\left(\hat R-2\Lambda\right)\right]+S_\text{rad},\label{gr-action}\eea where the effective cosmological constant $\Lambda=\lambda M^2_\text{pl}/8$ and we have taken into account that, for $\omega\neq 6$ and $\beta^2\sim 1$, the mass squared of the vector field $Q_\mu$: $m^2_Q=(6-\omega)M^2_\text{pl}/4\beta^2\propto M^2_\text{pl}$, so that the nonmetricity field decouples from the low-energy gravitational spectrum. This means that the effective geometrical structure of background space is Riemannian so that, in what follows we make the replacement: $\tilde W_4\rightarrow V_4$.

The action \eqref{gr-action} is just the Einstein-Hilbert (EH) action over Riemann $V_4$ space. The obtained representation, $${\cal G}_\text{GR}:\{{\cal M}_4\in V_4,g_{\mu\nu},\phi=M_\text{pl}\,|\,{\cal S}_\text{GR},\,{\cal C},...\},$$ where $ {\cal S}_\text{GR}=\{M^2_\text{pl},\,v_0M_\text{pl},\,\Lambda\}$, is called as GR gauge. In this specific gauge the manifest gauge symmetry of the theory \eqref{tot-action'} is lost.


Although GR itself is clearly not gauge invariant, in the present framework it is no more than one of the infinitely many equivalent gauges in the conformal equivalence class \eqref{c-class}: ${\cal K}=\{{\cal G}_1={\cal G}_\text{GR},{\cal G}_2,...,{\cal G}_j,...,{\cal G}_N|\;j\in\mathbb{N}\}.$ Hence, since gauge invariance is the underlying symmetry behind the class ${\cal K}$, GR is part of a bigger gauge invariant theory.


\subsubsection{The many GR worlds}


Let us discuss how the many-worlds picture arises in the simplest case: the GR gauge. Since, in order to fix the GR gauge, the choice of a constant value of the scalar field $\phi$ is arbitrary, depending of the chosen constant value of the scalar field, one has (in principle) an infinite set of GR copies with different values of the Planck mass, of the masses of the SM fields and of the cosmological constant, among others. The GR gauge represents itself a subclass ${\cal K}_\text{GR}$ within the conformal class ${\cal K}$ \eqref{c-class}, which comprises an infinite number of GR copies: $${\cal K}_\text{GR}=\left\{{\cal G}^0_\text{GR}, {\cal G}^1_\text{GR}, {\cal G}^2_\text{GR},...,{\cal G}^k_\text{GR},...\right\},$$ where $k=0,1,2,3...,N$ ($N\rightarrow\infty$) and the general element of the GR gauge can be expressed as, $${\cal G}^k_\text{GR}:\{{\cal M}_4\in V_4,\,g_{\mu\nu},\,\phi_{0k}\,|\,{\cal S}^\text{GR}_k,\,{\cal C},...\},$$ where the different constants $\phi_{0k}\in\mathbb R$ generate different sets of physical constants: $${\cal S}^\text{GR}_k=\left\{M^2_{\text{pl},k}=\phi^2_{0k},\,v_k=v_0\phi_{0k},\,\Lambda_k=\lambda\phi^2_{0k}/8\right\}.$$  

In this many-worlds approach to GR the experiment allows determining which one of the infinitely many GR gauges is the one which better describes our universe through associating experimental values to the measured quantities, in particular to the constants of nature.



\section{The many worlds: a cosmological example}\label{sect-example} 


For further illustration of the many-worlds approach to gauge symmetry, let us consider a cosmological example. Let us write the independent EOM in Eq. \eqref{eff-eom} -- these will be the Einstein's equations -- in terms of the Friedmann-Robertson-Walker (FRW) metric with flat spatial sections (in what follows, for simplicity, we omit the term $\propto\lambda\phi^4$),

\bea ds^2=-dt^2+a^2(t)\delta_{ij}dx^i dx^j,\label{frw-metric}\eea where $t$ is the cosmic time and $a(t)$ is the dimensionless scale factor. We get that,\footnote{There are not other independent equations. For instance, the Raychaudhuri equation is obtained by deriving \eqref{fried-eq} with respect to the cosmic time and taking into account the continuity equation. The equation for the scalar field coincides with the trace of \eqref{eff-eom}, so that it is not an independent equation.}

\bea 3\left(H+\frac{\dot\phi}{\phi}\right)^2=\frac{1}{\phi^2}\rho_\text{rad},\label{fried-eq}\eea where the radiation SET \eqref{rad-set} has been considered. The continuity equation $\hat\nabla^\mu T^\text{rad}_{\mu\nu}=0$, leads to $\dot\rho_\text{rad}+4H\rho_\text{rad}=0$, whose straightforward integration yields: $\rho_\text{rad}=\rho_0 a^{-4}$, where $\rho_0$ is an integration constant. If introduce the gauge invariant variable $\chi\equiv a\phi$ and replace the cosmic time by the gauge invariant conformal time: $d\tau=a^{-1}dt$, equation \eqref{fried-eq} can be given the form of a very simple gauge invariant equation:

\bea \chi'=\sqrt{\rho_0},\label{ginv-fried-eq}\eea where the tilde means derivative with respect to the conformal time $\tau$. This equation can be integrated in quadratures to get the following gauge invariant expression:

\bea a(\tau)\phi(\tau)=\sqrt{\rho_0}(\tau-\tau_0),\label{sol-ginv-eq}\eea where $\tau_0$ is an integration constant. Different choices of the function $\phi(\tau)$ fix different gauges. For brevity let us consider only three representative of them:

\begin{itemize}

\item The GR gauge where $\phi=\phi_0=$const.\footnote{Recall that there are possible infinitely many different choices of the constant $\phi_0$, so that there can be infinitely many copies of GR in the GR gauge.} In this case $a(\tau)=a_0(\tau-\tau_0),$ where $a_0=\sqrt{\rho_0}/\phi_0$. In terms of the cosmic time we have that $a(t)=\sqrt{2a_0}(t-t_0)^{1/2},$ where $t_0$ is another integration constant.

\item de Sitter gauge, where $$H=H_0\;\Rightarrow\;\frac{a'}{a^2}=H_0,$$ or, after integration: 

\bea a(\tau)=H^{-1}_0(\tau_0-\tau)^{-1}\Rightarrow a(t)=a_0\,e^{H_0t},\nonumber\eea where $a_0=\exp{(-H_0t_0)}/H_0$. In this case the squared gauge scalar evolves as,

\bea \phi^2(\tau)=H^2_0\rho_0\left(\tau-\tau_0\right)^4\Rightarrow\phi^2(t)=\phi^2_0\,e^{-4H_0t},\nonumber\eea where $\phi_0=\sqrt{\rho_0}a^2_0/H_0$.

\item Third (unphysical) gauge where contraction of the universe takes place instead of expansion. We assume $a(t)=t^{-n}$ where $n$ is a positive real number. In terms of the conformal time this amounts to: $a(\tau)=a_0(\tau-\tau_0)^{-\frac{n}{n+1}},$ and $\phi^2(\tau)=\phi^2_0(\tau-\tau_0)^\frac{2(2n+1)}{n+1}$ $\Rightarrow$ $\phi^2(t)=\bar\phi^2_0\,t^{2(2n+1)},$ where $a_0=(n+1)^{-n/(n+1)}$, $\phi_0=\sqrt{\rho_0}/a_0$ and $\bar\phi_0=\sqrt{\rho_0}/n+1$.

\end{itemize} 

The listed gauges can be expressed in the following general form:\footnote{For compactness here we omit explicit writing of the sets of constants of nature ${\cal S}(\tau)=\{M^2_\text{pl}(\tau),v(\tau),\Lambda(\tau)\}$ and ${\cal C}=\{\hbar,\,c,\,e,...\}$.} ${\cal G}:\left\{{\cal M}_4\in V_4,a(\tau),\vphi(\tau)\right\}$. More specifically these can be expressed as follows,

\bea &&{\cal G}_\text{GR}:\left\{{\cal M}_4\in V_4,\;a(\tau)=a_0(\tau-\tau_0),\;\phi(\tau)=\phi_0\right\},\nonumber\\
&&{\cal G}_\text{dS}:\left\{{\cal M}_4\in V_4,\;a(\tau)=H^{-1}_0(\tau_0-\tau)^{-1},\right.\nonumber\\
&&\;\;\;\;\;\;\;\;\;\;\;\;\;\;\;\;\;\left.\phi(\tau)=\sqrt{\rho_0}H_0\rho_0(\tau-\tau_0)^2\right\},\nonumber\\
&&{\cal G}_\text{third}:\left\{{\cal M}_4\in V_4,a(\tau)=a_0(\tau-\tau_0)^{-\frac{n}{n+1}},\right.\nonumber\\
&&\;\;\;\;\;\;\;\;\;\;\;\;\;\;\;\;\;\left.\phi(\tau)=\phi_0(\tau-\tau_0)^\frac{2n+1}{n+1}\right\}.\label{3-gauges}\eea These gauges are linked by gauge transformations \eqref{gauge-t}:

\bea a^2(\tau)\rightarrow\Omega^2(\tau)a^2(\tau),\;\phi^2(\tau)\rightarrow\Omega^{-2}(\tau)\phi^2(\tau).\label{conf-t}\eea For instance, the first two gauges are linked by these transformations with conformal factor $\Omega^2=a^2_0H^2_0(\tau-\tau_0)^4,$ while for the first and third gauges, $\Omega^2=(\tau-\tau_0)^\frac{2(2n+1)}{n+1}.$ All three gauges are associated with radiation domination since only radiation can be considered in \eqref{eff-eom}.

While the GR gauge describes a stage with decelerated expansion, the de Sitter gauge describes an inflationary period of the cosmic expansion and the third gauge represents a contracting universe. These different cosmological behaviors may be observationally differentiated. In particular, the gravitational constant measured in Cavendish-type experiments $G_N$, shows a different dynamical behavior in each gauge:

\bea &&8\pi G_N^\text{GR}=\phi^{-2}_0,\;8\pi G_N^\text{dS}=\phi^{-2}_0e^{4H_0t},\nonumber\\
&&8\pi G_N^\text{third}=\bar\phi^{-2}_0t^{-2(2n+1)},\nonumber\eea correspondingly. Therefore, assuming that one of these behaviors may correctly explain the radiation dominated cosmic dynamics, observations are able to select the corresponding gauge.\footnote{From the start there is a gauge which we know does not meet the observational data: the one which describes cosmic contraction.}



\section{Gradient nonmetricity}\label{sect-wig}  


In former sections it has been shown that gauge symmetry must be broken down before, or at least simultaneously, with $SU(2)\times U(1)$ symmetry. I. e., Weyl gauge symmetry does not survives after EW symmetry breaking. This is due to the fact that only massless fields couple to gravity in gauge invariant theories of gravity of class \eqref{eff-lag}. However, in Weyl integrable geometry (WIG) space, which we denote here as $\tilde W_4^\text{int}$, since the nonmetricity vector amounts to a gradient of a scalar field, we have an opportunity to improve the above issue. 

This can be done by lifting the gauge scalar field $\phi$ to the category of a geometric field. In other words, we assume that the nonmetricity of WIG space is given by, 

\bea \nabla_\alpha g_{\mu\nu}=-2\frac{\der_\alpha\phi}{\phi}g_{\mu\nu},\label{wig-nm}\eea i. e. that the nonmetricity vector $Q_\mu=2\der_\mu\phi/\phi$. Under this assumption we have that $\phi^2R=\phi^2\hat R-6\phi\hat\nabla^2\phi$ or, equivalently: $\phi^2 R=\phi^2\hat R+6(\der\phi)^2-6\hat\nabla^\mu\left(\phi\der_\mu\phi\right),$ where the last term amounts to a boundary term that can be omitted. The action of gauge invariant gravity in $\tilde W_4^\text{int}$ space reads

\bea &&S^\text{wig}_g=\frac{1}{2}\int d^4x\sqrt{-g}\phi^2R\nonumber\\
&&\;\;\;\;\;\;\;\;=\frac{1}{2}\int d^4x\sqrt{-g}\left[\phi^2\hat R+6(\der\phi)^2\right].\label{wig-action}\eea 

The most interesting property of the above action is that matter fields, whether massless or with the mass, couple to gravity without breaking the gauge symmetry. Consider the gauge invariant action over $\tilde W_4^\text{int}$:

\bea S^\text{wig}_\text{tot}=\frac{1}{2}\int d^4x\sqrt{-g}\left[\phi^2\hat R+6(\der\phi)^2+2{\cal L}_m\right],\label{wig-tot-action}\eea where ${\cal L}_m$ is the Lagrangian of the matter fields. Consistent coupling of arbitrary matter fields is possible thanks to the property that in WIG space variation of the metric is not independent of variation of the geometric scalar field $\phi$, since due to gradient nonmetricity law \eqref{wig-nm} one has that (see, for instance, Eq. (3) of Ref. \cite{jackiw-2015},)

\bea \delta g_{\mu\nu}=-2\frac{\delta\phi}{\phi}g_{\mu\nu},\;\delta g^{\mu\nu}=2\frac{\delta\phi}{\phi}g^{\mu\nu}.\label{wig-var}\eea This means, for instance, that variation of the overall Lagrangian in \eqref{wig-tot-action},

\bea &&{\cal L}_\text{tot}=\frac{1}{2}\left[\phi^2\hat R+6(\der\phi)^2+2{\cal L}_m\right],\label{wig-tot-lag}\\
&&\delta\left(\sqrt{-g}{\cal L}_\text{tot}\right)=\frac{\der\left(\sqrt{-g}{\cal L}_\text{tot}\right)}{\der g^{\mu\nu}}\delta g^{\mu\nu}\nonumber\\&&\;\;\;\;\;\;\;\;\;\;\;\;\;\;\;\;\;\;\;\;\;=2\frac{\der\left(\sqrt{-g}{\cal L}_\text{tot}\right)}{\der g^{\mu\nu}}g^{\mu\nu}\frac{\delta\phi}{\phi}.\label{var-wig-tot-lag}\eea Hence, since

\bea &&\frac{\der\left(\sqrt{-g}{\cal L}_\text{tot}\right)}{\der g^{\mu\nu}}=\frac{\sqrt{-g}}{2}\left[\phi^2G_{\mu\nu}-T^{(m)}_{\mu\nu}\right]\nonumber\\
&&\;\;\;\;\;\;\;\;=\frac{\sqrt{-g}}{2}\left\{\phi^2\hat G_{\mu\nu}+6\left[\der_\mu\phi\der_\nu\phi-\frac{1}{2}g_{\mu\nu}(\der\phi)^2\right]\right.\nonumber\\
&&\;\;\;\;\;\;\;\;\;\;\;\;\;\;\;\;\;\;\;\;\;\;\;\;\left.-\left(\hat\nabla_\mu\hat\nabla_\nu-g_{\mu\nu}\hat\nabla^2\right)\phi^2-T^{(m)}_{\mu\nu}\right\},\nonumber\eea variation of the action \eqref{wig-tot-action} with respect to the metric yields the Einstein's EOM,

\bea &&G_{\mu\nu}=\frac{1}{\phi^2}T^{(m)}_{\mu\nu}\;\Leftrightarrow\nonumber\\
&&\hat G_{\mu\nu}-\frac{1}{\phi^2}\left(\hat\nabla_\mu\hat\nabla_\nu-g_{\mu\nu}\hat\nabla^2\right)\phi^2\nonumber\\
&&\;\;\;\;\;\;\;+\frac{6}{\phi^2}\left[\der_\mu\phi\der_\nu\phi-\frac{1}{2}g_{\mu\nu}(\der\phi)^2\right]=\frac{1}{\phi^2}T^{(m)}_{\mu\nu},\label{wig-einst-eom}\eea where we have taken into account that the Einstein's tensor of $\tilde W^\text{int}_4$ space can be written in terms of LC (Riemannian) quantities according to 

\bea &&G_{\mu\nu}=\hat G_{\mu\nu}-\frac{1}{\phi^2}\left(\hat\nabla_\mu\hat\nabla_\nu-g_{\mu\nu}\hat\nabla^2\right)\phi^2\nonumber\\
&&\;\;\;\;\;\;\;\;\;\;\;\;\;\;\;\;\;\;+\frac{6}{\phi^2}\left[\der_\mu\phi\der_\nu\phi-\frac{1}{2}g_{\mu\nu}(\der\phi)^2\right].\nonumber\eea Meanwhile, according to \eqref{var-wig-tot-lag},

\bea &&\delta_\phi\left(\sqrt{-g}{\cal L}_\text{tot}\right)=2\frac{\der\left(\sqrt{-g}{\cal L}_\text{tot}\right)}{\der g^{\mu\nu}}g^{\mu\nu}\frac{\delta\phi}{\phi}\nonumber\\
&&\;\;\;\;\;\;\;\;\;\;\;\;\;\;\;\;\;\;\;\;\;\;=-\sqrt{-g}\left[\phi^2R+T^{(m)}\right]\frac{\delta\phi}{\phi}.\nonumber\eea Hence, variation of \eqref{wig-tot-action} with respect to $\phi$ leads to:

\bea -R=-\hat R-6\frac{(\der\phi)^2}{\phi^2}+3\frac{\hat\nabla^2\phi^2}{\phi^2}=\frac{1}{\phi^2}T^{(m)},\label{wig-kg-eom}\eea which coincides with the trace of the Einstein's EOM \eqref{wig-einst-eom} without requiring vanishing SET trace. In consequence, the geometric gauge scalar $\phi$ is not a dynamical field: it can be chosen at will. Different choices lead to different gauges.


\subsection{Continuity equation}


In section \ref{sect-lemma-1} we have demonstrated that the standard continuity equation \eqref{gr-cons-eq} in background Riemann space $V_4$: $\hat\nabla^\mu T^{(m)}_{\mu\nu}=0$, takes place in the class of gauge invariant theories \eqref{g-lag} over Weyl space $\tilde W_4$. This means that radiation and massless SM fields: the only matter fields which couple to gravity in this framework, follow null geodesics of Reimann space. In consequence we may drop out the nonmetricity vector, i. e. we may replace $\tilde W_4\rightarrow V_4.$

In the present case where a gauge invariant theory over WIG space -- distinguished by gradient nonmetricity \eqref{wig-nm} -- is considered, the continuity equation can be derived in the following way. Let us take the LC covariant divergence of the quantity $\phi^2G_{\mu\nu}$. According to the EOM \eqref{wig-einst-eom} we get that,

\bea &&\hat\nabla^\mu\left(\phi^2G_{\mu\nu}\right)=\hat\nabla^\mu\phi^2\hat G_{\mu\nu}-(\hat\nabla^2\hat\nabla_\mu-\hat\nabla_\mu\hat\nabla^2)\phi^2\nonumber\\
&&\;\;\;\;\;\;\;\;\;\;\;\;\;\;\;\;\;\;\;\;\;\;\;\;+\frac{\hat\nabla_\nu\phi^2}{2}\left[3\frac{\hat\nabla^2\phi^2}{\phi^2}-6\frac{(\der\phi)^2}{\phi^2}\right],\nonumber\eea or, if consider the equation \eqref{a-usef-1} with the replacement $\vphi\rightarrow\phi^2$, we obtain that

\bea &&\hat\nabla^\mu\left(\phi^2G_{\mu\nu}\right)=\frac{\hat\nabla_\nu\phi^2}{2}\left[-\hat R-6\frac{(\der\phi)^2}{\phi^2}+3\frac{\hat\nabla^2\phi^2}{\phi^2}\right].\nonumber\eea Hence, if in this equation substitute \eqref{wig-kg-eom} and take into account the EOM \eqref{wig-einst-eom}: $\phi^2G_{\mu\nu}=T^{(m)}_{\mu\nu}$, we finally obtain the following continuity equation (compare with Eq. \eqref{gr-cons-eq}):

\bea \hat\nabla^\mu T^{(m)}_{\mu\nu}=\frac{\hat\nabla_\nu\phi^2}{2\phi^2}T^{(m)}=\frac{\hat\nabla_\nu\phi}{\phi}T^{(m)}.\label{wig-cont-1}\eea This equation means that matter fields with traceless SET: $T^{(m)}=0$, follow null geodesics of Riemann space $V_4$, meanwhile, SM matter fields with nonvanishing $T^{(m)}\neq 0$, follow timelike geodesics of WIG space $\tilde W^\text{int}_4$ instead. 

Given that in the theory \eqref{wig-tot-action} all of SM fields couple to gravity, no matter whether massless or with the mass, gauge symmetry in this theoretical framework may have impact in the phenomenology after EW symmetry breaking. Investigation of this impact is the subject of \cite{quiros-arxiv-2023}.



\section{Discussion and conclusion}\label{sect-discu}


In this paper we have investigated gauge invariant gravitational Lagrangians of the general form \eqref{g-lag}. This Lagrangian leads to the independent second-order EOM \eqref{S-einst-eom} and \eqref{max-eom}, despite that it contains a quadratic curvature term of the peculiar form: $Q^2=R_{(\alpha\beta)\mu\nu}R^{(\alpha\beta)\mu\nu}=R_{[\mu\nu]}R^{[\mu\nu]}.$ One of the main results of this paper is comprised in lemma \ref{lemma-1}: ``In gauge invariant gravitational theories of class \eqref{g-lag} (i) only matter fields with traceless SET couple to gravity and (ii) these follow geodesics of Riemann geometry,'' and in its corollary \ref{coro-l1}: ``Massless fields -- the only matter fields which satisfy the EOM of theories of class \eqref{g-lag} -- do not interact with the nonmetricity vector $Q_\mu$.'' The lemma (together with its proof and the resulting corollary) generalizes previous works where it is shown that the nonmetricity vector does not interact with massless fermions \cite{cheng-1988, cheng-arxiv}. 

Consideration of other quadratic terms, like $R^2$, leads to the particular form of Lagrangian \eqref{g-lag}, where $\omega=0$ (no kinetic term for the scalar field,) as it is shown in \cite{ghilen-1}. In this bibliographic reference a well known method \cite{tourrenc-1983, wands-cqg-1994}: to replace the quadratic term by a linear term multiplied by a non dynamical scalar $R^2\rightarrow-2\phi^2R-\phi^4$, is used to ``linearize'' the gravitational Lagrangian. 

Addition of a quadratic term of the form,\footnote{This theory, which has been developed in \cite{mannheim-1989, mannheim-2006}, has severe problems. According to \cite{ccorr-stelle-1978} (see also \cite{hindawi-1996}), to lowest order, fluctuations of the metric around flat space in a theory with quadratic action of the form: $$S_\text{st}=\frac{1}{2}\int d^4x\sqrt{-g}\left[\hat R+\frac{1}{6m_0^2}\hat R^2-\frac{1}{2m_2^2}\hat C^2\right],$$ lead to a perturbations spectrum which, in addition to the graviton, contains a scalar field with mass $m_0$ and a spin-two field with mass $m_2$. This quadratic theory is renormalizable but nonunitary. The theory \eqref{mannheim-action}, in contrast, contains only the ghost-like spin-two field and has no graviton in its spectrum. This rules out this theory as a phenomenologically viable description of low curvature gravitational phenomena.}

\bea &&S_W=\alpha\int d^4x\sqrt{-g}\hat C^2,\label{mannheim-action}\eea where $\alpha$ is a dimensionless constant, $\hat C^2\equiv\hat C_{\mu\lambda\nu\sigma}\hat C^{\mu\lambda\nu\sigma}$, and $\hat C_{\mu\lambda\nu\sigma}$ is the Weyl tensor of Riemann space, does not modify the results of lemma \ref{lemma-1} and of its corollary either. Actually, in this case the derived EOM reads \cite{mannheim-1989}:

\bea \hat W^{(2)}_{\mu\nu}-\frac{1}{3}\hat W^{(1)}_{\mu\nu}=\frac{1}{4\alpha}T^\text{mat}_{\mu\nu},\label{mannheim-eom}\eea where 

\bea &&\hat W^{(1)}_{\mu\nu}=2g_{\mu\nu}\hat\nabla^2\hat R-2\hat\nabla_\mu\hat\nabla_\nu\hat R-2\hat R\hat R_{\mu\nu}+\frac{1}{2}g_{\mu\nu}\hat R^2,\nonumber\\
&&\hat W^{(2)}_{\mu\nu}=\frac{1}{2}g_{\mu\nu}\hat\nabla^2\hat R+\hat\nabla^2\hat R_{\mu\nu}-2\hat\nabla_{(\mu}\hat\nabla_\lambda\hat R^{\;\;\lambda}_{\nu)}\nonumber\\
&&\;\;\;\;\;\;\;\;\;\;\;\;-2\hat R_{\mu\lambda}\hat R^{\;\;\lambda}_\nu+\frac{1}{2}g_{\mu\nu}\hat R_{\lambda\sigma}\hat R^{\lambda\sigma}.\nonumber\eea The trace of Eq. \eqref{mannheim-eom} yields,

\bea \frac{1}{4\alpha}T^\text{mat}=\hat\nabla^2\hat R-2\hat\nabla^\mu\hat\nabla^\nu\hat R_{\mu\nu}=-2\hat\nabla^\mu\hat\nabla^\nu\hat G_{\mu\nu},\nonumber\eea which exactly vanishes thanks to the Bianchy identity $\hat\nabla^\mu\hat G_{\mu\nu}=0$. Hence, only radiation couples to gravity in this theory. Despite that \eqref{mannheim-action} has been proposed as a possible explanation to the dark matter issue, quite the contrary effect is obtained since the DM does not interact with radiation. In a cosmological context this theory could describe the radiation dominated epoch of the cosmic evolution exclusively. But the matter dominated stage, where the formation of cosmic structure happens and where the dark matter plays the most important part, requires of a different theory that should replace \eqref{mannheim-action}. Hence, dark matter can not be explained in the present setup as incorrectly claimed in \cite{mannheim-1989, mannheim-2006}.

In a similar fashion, the theory developed in \cite{ghilen-1, ghilen-2023, ghilen-2, ghilen-3, ghilen-4} which corresponds to the particular case when in \eqref{g-lag} the coupling vanishes: $\omega=0$, has been ``seemingly'' established in \cite{harko-prd-2023} as a basis for an alternative explanation of the DM. We underline the word ``seemingly'' because explaining the DM in the class of theories \eqref{g-lag} is forbidden by lemma \ref{lemma-1} and its corollary. In the mentioned reference the authors look for static, spherically symmetric, vacuum solutions to the Einstein's EOM \eqref{einst-eom} with $\omega=0$ (also with $\lambda=0$.) Then they investigate the physical properties of the stable circular timelike geodesic orbits of massive test particles in static, spherically symmetric, vacuum space $\tilde W_4$. This is misleading since only massless fields can couple to gravity in this gauge invariant theory, as stated in lemma \ref{lemma-1}. A feasible physical explanation of lemma \ref{lemma-1} in this case, can be based on Proca equation \eqref{max-eom} with $\omega=0$ (see also the Stuckelberg-type Lagrangian ${\cal L}_S$ \eqref{stueck-grav-lag}). The Weyl vector field $Q_\mu$ has an effective mass squared $m^2_Q(x)=3M^2_\text{pl}(x)/2\beta^2=3\phi^2(x)/2\beta^2$, where the effective Planck mass $M_\text{pl}(x)$ sets the grand unification scale point by point in space. Hence, the nonmetricity vector is a short range field with range $\sim M^{-1}_\text{pl}$. This means it is strongly screened so that it does not modify the motion of test fields in any appreciable way. As a matter of fact the $Q_\mu$-s decouple from the low-energy spectrum in general.

Hence, what do the authors of \cite{harko-prd-2023} have really done? In order to find exact solutions they assumed that the nonmetricity vector has only a nonvanishing radial component $Q_\mu=(0,w_r,0,0)$. This implies that $Q_{\mu\nu}=0$ $\Rightarrow Q^2=0$. If set $Q_{\mu\nu}=0$ in \eqref{max-eom}, it follows that $Q_\mu=\der_\mu\phi^2/\phi^2$, i. e., the geometric structure of background space is Weyl integrable geometry with nonmetricity law \eqref{wig-nm}, instead of just Weyl geometry. Hence, in the setup investigated in \cite{harko-prd-2023} one has to make the replacement $\tilde W^\text{int}_4\rightarrow\tilde W_4$. Besides, since $Q^2=0$ and $Q_\mu=\der_\mu\phi^2/\phi^2$, then the Stuckelberg-type Lagrangian \eqref{stueck-grav-lag} vanishes as well. We are led with the effective gravitational Lagrangian \eqref{eff-lag}: $${\cal L}^\text{eff}_g=\frac{1}{2}\left[\phi^2\hat R+6(\der\phi)^2-\frac{\lambda}{4}\phi^4\right],$$ which up to the irrelevant term $\propto\phi^4$ coincides with the gravitational Lagrangian in \eqref{wig-action}. Hence, the model which is investigated in \cite{harko-prd-2023} as the basis for the explanation of the DM is not the model assumed by the authors, but the one studied in section \ref{sect-wig}, which is based in background space with gradient nonmetricity \eqref{wig-nm}, so that it evades the lemma \ref{lemma-1} and its corollary.


Perhaps the main lesson to be learn from the present investigation is that the class of gauge invariant theories given by the Lagrangian \eqref{g-lag} can have an impact only in the dynamics during the radiation epoch, before $SU(2)\times U(1)$ symmetry breaking takes place. Hence neither the dark matter nor the dark energy can be linked with the nonmetricity vector (Weyl gauge vector.) This result can be easily extended to the case when in \eqref{g-lag} replace the scalar field $\phi$ by a multicomponent (complex) scalar $\vphi$, such that, $\phi^2\rightarrow|\vphi|^2=\vphi^\dag\vphi,$ $(\der^*\phi)^2\rightarrow|\der^*\vphi|^2$, $$|\der^*\vphi|^2=|\der\vphi|^2-\frac{1}{2}\der_\mu|\vphi|^2Q^\mu+\frac{1}{4}|\vphi|^2Q_\mu Q^\mu.$$ A consequence of this result is that, for instance, the theoretical framework proposed in \cite{cheng-1988} can not explain the DM as incorrectly suggested in that paper (see also the subsequent \cite{cheng-arxiv}).


A controversial aspect of the present study can be related with our approach to gauge symmetry. We have approached to the physical and geometrical interpretation of gauge invariance from a different perspective, where a gauge choice has physical consequences, so that it is subject to experimental check. According to our approach gauge freedom can be associated with a physical picture resembling the many-worlds interpretation of quantum physics. The gauge scalar $\phi$ does not obey any specific EOM so that it may be fixed at will. This means that in equations \eqref{phi-eom}, \eqref{max-eom} and \eqref{einst-mat-eom} we may choose any function $\vphi(x)$ we want. The result will be a specific theory associated with this choice of a gauge. Hence, each gauge represents a whole theory of gravity, which is characterized by a specific behavior in spacetime of several fundamental ``constants,'' the mass of the SM particles, etc. An outstanding gauge in this theoretical framework is the so called GR gauge, which is a (in principle infinite) set of copies of GR theory, specified by the choice $\phi=\phi_{0k}$ ($k=1,2,...,N$), where the $\phi_{0k}$ are different constants. In this gauge the gravitational laws look exactly the same, so that each member in the GR gauge differs from any other in the values of the measured Newton's constant $8\pi G_{N,k}=M^{-2}_{\text{pl},k}$ and of the EW mass parameter $v^2_{0k}$, among others. In our framework general relativity is just a subclass of a bigger theory. Manifest gauge symmetry is lost once a specific gauge has been chosen. This is why GR seems to evade this symmetry. Yet, it is a residual symmetry since any specific gauge is related with any other gauge through the gauge transformations \eqref{gauge-t}. 

The present classic gravitational version of the many-worlds interpretation of quantum physics is interesting because it provides a different perspective on the relation between theory and experiment: The experiment allows to determine which one of the infinitely many gauges is the one which better describes our universe through associating experimental values to the constants of nature.


We conclude that, despite being decoupled from the low-energy gravitational spectrum, vectorial nonmetricity and gauge symmetry may have led their footprints in the quantum era. When in the above theory we replace vector by gradient nonmetricity, the resulting gauge invariant theoretical framework, which is given by Lagrangian \eqref{wig-tot-lag} and the derived EOM \eqref{wig-einst-eom}, is the only possibility left to us by nature to search for the classical phenomenological and observational consequences of gauge symmetry. This theory is investigated in a separate publication \cite{quiros-arxiv-2023}.


{\bf Acknowledgments.} The author acknowledges FORDECYT-PRONACES-CONACYT for support of the present research under grant CF-MG-2558591.




\end{document}